\documentstyle[epsfig, twocolumn]{esapub} 
 
\begin{document} 
 
\def\note #1]{{\bf #1]}}
\def\figdir{./fig}
\def\ie{{\it i.e.}}
\def\muHz{\,\mu{\rm Hz}}
\def\picplace#1{\vspace{#1}}
\def\ddt#1{{\partial #1\over\partial t}}
\def\ddz#1{{\partial #1\over \partial z}}
\def\DDt#1{{{\rm D} #1\over {\rm D} t}}
\def\DD{{\rm D}}
\def\dd{{\rm d}}
\def\vu{\vec{u}}
\def\vxi{\vec{\xi}}
\def\pg{{p_{\rm g}}}
\def\pt{{p_{\rm t}}}
\def\rhobar{{\bar\rho}}
\def\cph{c_{\rm ph}}
\def\ubar{{\bar u}}
\def\uzprime{u_z^\prime}
\def\pbar{{\bar p}}
\def\pgbar{{\bar p}_{\rm g}}
\def\uzbar{{\bar u_z}}
\def\bra{\langle}
\def\ket{\rangle}
\def\Sect#1{Section \ref{#1}}
\def\Eq#1{Eq.\ (\ref{#1})}
\def\Fig#1{Fig.\ \ref{#1}}
\def\apj{{\rm ApJ}}
\def\s{\, {\rm s}}
\def\km{\, {\rm km}} 
 
%% Do not remove the following six lines: 
\setlength{\parindent}{0pt} 
\setlength{\parskip}{ 10pt plus 1pt minus 1pt} 
\setlength{\hoffset}{-1.5truecm} 
\setlength{\textwidth}{ 17.1truecm } 
\setlength{\columnsep}{1truecm } 
\setlength{\columnseprule}{0pt} 
\setlength{\headheight}{12pt} 
\setlength{\headsep}{20pt} 
\pagestyle{esapubheadings} 
 
%% Title - should be in capitals: 
\title{\bf P-MODE LINE-SHAPES IN THEORY AND PRACTICE} 
%% If the author list spans more than one line then the {\bf (bold 
%% font)} command must be inserted for each line 
\author{{\bf C.S. Rosenthal}
\vspace{2mm} \\ 
High Altitude Observatory/NCAR, P.O. Box 3000, Boulder, CO 80307-3000 \\ 
} 
 
\maketitle 
 
\begin{abstract} 

The asymmetries of solar p-mode lines provide important diagnostic clues as
to the nature and location of the source of mode excitation. At the same time,
they can also be responsible for systematic errors in the determination of
mode frequencies.

I present a matched wave asymptotic analysis which leads to an expression for
the spectrum valid for arbitrary stratifications and arbitrary source types.
From this it is possible to deduce the correct asymmetric line profile to use
in fitting the spectra, along with the physical interpretation of the fitted
parameters.

I apply this procedure to low-degree MDI doppler data from which I am able to
deduce corrections to the mode frequencies. A preliminary comparison of the
asymmetry parameter with numerical calculations yields an estimate of the
source depth. \vspace {5pt} \\ 
 
%% Do not remove the previous commands. Your abstract should  
%% end with \vspace {5pt} \\   

%% Please insert your keywords here. 
  Key~words: solar oscillations; line asymmetry; excitation.

\end{abstract} 
 
\section{INTRODUCTION} 
 
The discovery of the asymmetry of solar p-mode lines dates back to 
\cite*{Duvall+93}, but it is only with recent SOHO and GONG data that
we can now hope to see the asymmetry clearly in individual lines. 
Consequently it is only recently that we have begun to worry about
possible systematic errors in frequency determinations which might arise
as a result of ignoring the asymmetries. Essentially there are three
questions in which we are interested :- 1) What is the correct line profile
to use in fitting asymmetric p-mode lines? 2) What can we learn about the
source of excitation from the asymmetries?, and 3) Why do the asymmetries appear
oppositely signed in velocity and intensity data? In this paper I will
concentrate on (1) and (2) and comment only briefly on (3).

There have been numerous models of the asymmetry 
(\cite{Gabriel92,Gabriel95,Abrams+Kumar96}) from which it is clear that solar
oscillations driven near the upper turning point will inevitably have asymmetric
line-profiles whose properties will be sensitive to the location and type of
the source.
The discussion of \cite*{Rast+Bogdan98}
is particularly illuminating because they show that the minima or troughs in the
spectrum are (at least in some circumstances) eigenfrequencies of a solar model
truncated from above at the location of the source, and calculated with an upper
boundary condition determined by the source type. The relative location of the 
peaks and troughs is then responsible for the asymmetry of the lines.
This suggests that, if one
could determine the locations of these troughs, one could analyse
the frequencies of these minima using standard helioseismic techniques. For
example, the trough-frequencies must satisfy a Duvall Law and a variational 
principle. This conceptual framework provides the basis for much of the following
discussion.

\section{GENERAL ASYMPTOTIC THEORY} 
 
The general asymptotic theory is described in \cite*{Rosenthal98} (in what follows,
R98). The 
basic problem is the solution of the oscillation equation for a plane-parallel
stratified layer in the form 
\begin{equation}
{\dd^2\Psi\over \dd z^2}+K^2\Psi=A_{\omega,k}(z)
\label{general_eqn}
\end{equation}
where the local vertical wavenumber is given by
\begin{equation}
K^2={1\over c^2}\left[ \omega^2-\omega_c^2-c^2k^2\left(1-{N^2\over\omega^2}\right)\right]
\end{equation}
and the source is assumed to have the form 
\begin{equation}
A_{\omega,k}(z)=a_0|K(z_0)|^{1\over 2}\delta(z-z_0)+a_1|K(z_0)|^{-{1\over
2}}\delta^\prime(z-z_0).
\label{local}
\end{equation}
Note that the solution with $a_1\!=\!0$ is the Green's Function and can therefore be
used to generate the solution for a truly arbitrary source. For brevity I here
consider only the low-frequency case in which the equation has two turning points
and the source lies wholly above the upper turning point. The intermediate and
high frequency cases, and the solution for an arbitrary source variation with depth, are
discussed in R98. In the low-frequency case the Fourier
spectrum in the region above the source is found to be
\begin{equation}
F={1\over 2}(-a_0-a_1)+{1\over 4}(-a_0+a_1)e^{-2\Delta}\tan\phi
\label{spec1}
\end{equation}
where
\begin{equation}
\phi\equiv\int_{Z_t}^{z_t}K\dd z.
\label{phi}
\end{equation}
and
\begin{equation}
\Delta\equiv\int_{z_0}^{Z_t} |K^2|^{1\over2} \dd z
\label{Delta}
\end{equation}
and $Z_t$ and $z_t$ are the lower and upper turning points, respectively.

\subsection{Properties of the Peaks and Troughs}

The spectrum diverges at frequencies satisfying
\begin{equation}
\int_{Z_t}^{z_t}K\dd z=\left(n-{1\over 2}\right)\pi.
\label{modes1}
\end{equation}
This is a well-known expression for the asymptotic frequencies of p modes and, as
one would expect, it does not depend on the source properties.

If the ratio ${ a_0+a_1/a_0-a_1}$ is real, for example if either $a_0$ or $a_1$ is zero,
then the spectrum also has zeroes whose location is given by
\begin{equation}
\int_{Z_t}^{z_t}K\dd z\approx\left(n-{1\over 2}\right)\pi+2e^{-2\Delta}{ a_0+a_1\over
a_0-a_1}.
\label{zeros1}
\end{equation}
If $\Delta$ is large, which is to say that the source lies well inside the upper
evanescent region, the troughs will lie close to the corresponding peaks and the asymmetry
will be large. The sign of the asymmetry clearly depends on the source type. A source 
containing only $a_0$ (a monopole source) will have a zero lying immediately above the mode 
frequency, and will
consequently have a negative asymmetry, as is seen in the MDI and GONG doppler data. 
For a dipole source the sign of asymmery is reversed.

If $\Delta$ and the source amplitudes are functions only of frequency
then equations (\ref{modes1}) and (\ref{zeros1}) are both in the form of Duvall Laws, 
$F_D\left({\omega/k}\right)=\pi({n+\alpha(\omega)/\omega })$ with the same Duvall function, $F_D$,
but differing phase functions $\alpha$:
\begin{equation}
\delta\alpha\equiv\alpha_t-\alpha_m= {2\over\pi}e^{-2\Delta}{ a_0+a_1\over a_0-a_1}
\label{dalpha}
\end{equation}
where $\alpha_t$ and $\alpha_m$ are the phase functions describing the troughs
and peaks respectively. Evidently the sign of $\delta\alpha$ is positive for
a pure monopole source and negative for a pure dipole, while its magnitude depends
quite strongly on the source location. The higher the source is placed, the smaller
the value of $\delta\alpha$, the closer
together the peaks and troughs, and the greater the asymmetry. I will therefore use
$\delta\alpha$ as a diagnostic for the source.

\subsubsection{A Solar-Cycle Dependence in the Asymmetry?}

Formally the WKB theory predicts that the mode frequencies depend on the strucure between
the turning points, while the trough frequencies depend on the structure between the
lower turning point and the source location. However, while the former statement is
only an approximation, the latter statement can actually be made exact. If the phase of
the source is constant across its depth, and the phase of the oscillation is constant
across the source, there will exist a family of nulls corresponding to frequencies at which
the amplitude of the solution is identically zero in the region above the source. The
location of these nulls is (obviously) strictly independent of the stratification in 
the region above the source, since the amplitude is identically zero there.
Thus any changes to the atmospheric structure as
a result, say, of solar-cycle effects, would change the mode frequencies but leave the
trough frequencies fixed. There therefore exists the possibility of a solar-cycle
dependence in the asymmetry. Any such effect is likely to be extremely small as the
asymmetry is significant only for low-frequency modes whose frequencies vary only very
slightly over the solar cycle.

\subsubsection{Intensity and Velocity Data}

Equation (\ref{spec1}) is the solution for the acoustic wave field.
\nocite{Nigam+98,Nigam+Kosovichev98,Roxburgh+Vorontsov97}Nigam et al (1998),
Nigam \& Kosovichev (1998) and Roxburgh \& Vorontsov (1997) all
explain the reversal of asymmetry between doppler and intensity measurements as
being due to the addition of correlated noise, by which they mean that the actual 
measured
signal, in at least one of the measurements, is contaminated by the source. This is
equivalent to adding terms proportional to $a_0$ and/or $a_1$ to the right hand side
of equation (\ref{spec1}). It is clearly therefore extremely difficult to tell the
difference between a change of source type and the addition of correlated noise. In
the data analysis below I will essentially assume that the doppler data 
represents the true acoustic wave field and that correlated noise is absent in the
doppler signal. At present it is not clear how that assumption should be validated
from the data.

\section{Application to Data}

\subsection{Asymmetric Line Profiles}

Equation (\ref{spec1}) suggests the following form for the power in the asymmetric
peaks:
\begin{eqnarray}
&\!\!\!\!\!P_1(\nu)&=\alpha_2+\alpha_3\times\cr
&&\left|1+\left(\alpha_4+i\alpha_5\right)\tan\left(
{\pi\over\delta}(\nu-\nu_0)+{\pi\over 2}+i\alpha_1\right)\right|^2\cr
\end{eqnarray}
which approximates to
\begin{equation}
P_1(\nu)\approx \alpha_2+\alpha_3{
(x-\alpha_4)^2+(\alpha_1-\alpha_5)^2
\over
x^2+\alpha_1^2}
\label{profile2}
\end{equation}
where
\begin{equation}
x\equiv {\pi\over\delta}(\nu-\nu_0).
\end{equation}

In practice I have found that $\alpha_5$, which is non-zero only if there is a
phase-variation across the source, is underconstrained by the data, so the final
preferred form for fitting is
\begin{equation}
P_2(\nu)=\alpha_2+\alpha_3+{\alpha_4^2\alpha_3\over \left(\nu-\nu_0\right)^2+\left( {\Gamma
\over 2}\right)^2}\displaystyle{\left[1-2\displaystyle{\nu-\nu_0\over\alpha_4}\right]}
\label{profile3}
\end{equation}
where the parameters to be fitted are $\alpha_2$, $\alpha_3$, $\alpha_4$, the eigenfrequency
$\nu_0$ and the full-width at half-maximum, $\Gamma$. The asymmetry is evidently due
entirely to $\alpha_4$. These various forms for the asymmetric profile are all broadly
similar to that proposed by \cite*{Nigam+98}. However, attention should be drawn to the
fact that the profiles proposed here are based on a general solution to the excitation
problem valid for an arbitrary stratification and source profile, and asymptotic forms
for the various parameters (in particular $\alpha_4$) can be easily deduced by comparison
with equation (\ref{spec1}), to yield a ready theoretical interpretation of the
fitted values. 

\subsection{Data Analysis}

The data analysed consists of 360 day time series of MDI doppler-velocity spherical-harmonic
coefficients for degrees $\ell\le 10$. (The gap-filled and detrended data were provided 
by Jesper Schou.) The decision to look at low-degree modes was based on the fact these are
less affected by leakage from modes of nearby degree. The remaining problem to be addressed
is therefore the m-leakage from modes of differing azimuthal order within the same
multiplet. For this preliminary analysis, the m-leakage issue was sidestepped by looking
only at sectoral modes for which the effect is negligible. Moreover, the effect of m-leakage
on the asymmetries of prograde and retrograde modes should be oppositely directed. Hence if
the fitted values of $\alpha_4$ agree between prograde and retrograde modes this provides
an {\it a posteriori}\/ justification for neglecting m-leakage.

The time series were divided up into ten equal months and FFT power spectra  calculated
for each month. The results were then averaged over the ten months. Fitted a-coefficients were
used to estimate mode frequencies and all peaks of a particular $m/l$ value in a given 
bin in degree and frequency were averaged to find a mean line profile. Figure 1 shows a
typical example for a low-frequency bin. The different modes have very similar line
profiles and show a clear negative asymmetry.

 \begin{figure}[!h] 
  \begin{center} 
    \leavevmode 
 \centerline{\epsfig{file=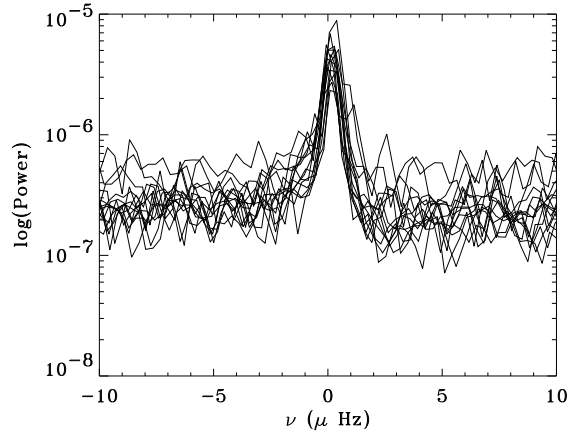,width=8.0cm}} 
%  \vspace{8.cm} 
  \end{center} 
  \caption{\em Power spectra of all retrograde sectoral modes in the bin $1\!\le\!\ell\!\le\!10$
  and $1.7\!\le\!\nu\!\le\!1.9${\rm mHz}, aligned using fitted a-coefficients.} 
%  \label{fig:residuals_S} 
\end{figure} 

The peaks were then averaged to obtain a mean line profile, which is then fit to 
the theoretically derived profile (\ref{profile3}). Because each peak is an average of
several modes over ten monthly spectra I assume that Gaussian statistics are valid and
the fitting was therefore carried out using a chi-squared minimisation with a weighting
proportional to the inverse-square of the power. Figure 2 shows the result of the fitting
for the same case as in Figure 1. Evidently the asymmetric profile provides a much better
fit to the data than the symmetric Lorentzian.

 \begin{figure}[!h] 
  \begin{center} 
    \leavevmode 
  \centerline{\epsfig{file=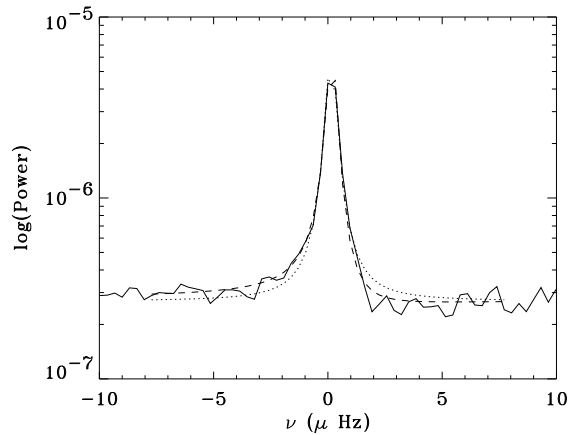,width=8.0cm}}
  \end{center} 
  \caption{\em Mean profile of the same modes as in Figure 1. The dotted curve is a fit of
  a symmetric Lorentzian profile and the dashed curve is a fit of the asymmetric profile,
  Eq. (\ref{profile3}).} 
%  \label{fig:residuals_S} 
\end{figure} 

\subsection{Frequency Corrections}

Figure 3 shows the difference in fitted frequencies between fits made using symmetric
and asymmetric profiles. For the sectoral modes, the difference is small at low frequencies,
where the line is so narrow that that the asymmetry makes little difference to the
fitted frequency. At higher frequencies, the lines are sufficiently symmetric that the
asymmetry coefficient is difficult to determine. The frequency error appears to decrease again
above 2.5mHz, but there is a large scatter. The maximum error is about $0.1\mu$Hz 
in the $2\!-\!2.5$mHz range.
There is no obvious systematic variation in the frequency error as a function of degree 
for these low-degree modes.
 \begin{figure}[!h] 
  \begin{center} 
    \leavevmode 
 \centerline{\epsfig{file=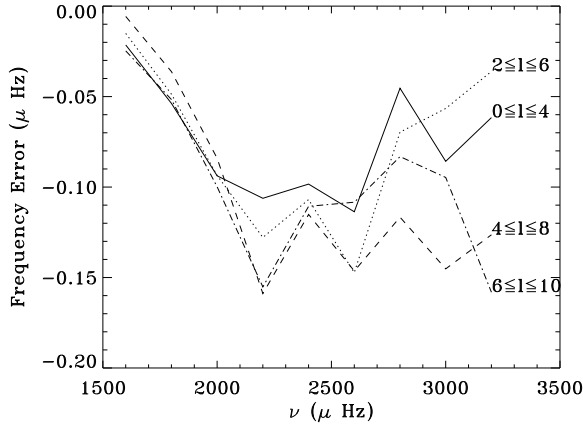,width=8.0cm}}
  \end{center} 
  \caption{\em Frequency difference in the sense (asymmetric fit) - (symmetric fit) for
  sectoral modes in the indicated $\ell$-bins. Results shown are averages of the results
  for prograde and retrograde modes.} 
%  \label{fig:residuals_S} 
\end{figure} 

\subsection{Asymmetry Parameter}

Figure 4 shows the asymmetry parameter $\alpha_4/2$ which is a measure of the 
separation of the peak and trough. The results for prograde and retrograde sectoral
modes are extremely similar (see also Figure 5), providing an {\em a posteriori}\/ justification for the
assumption that they are unaffected by m-leakage. Once again, there is no clear
systematic variation with degree.
 \begin{figure}[!h] 
  \begin{center} 
    \leavevmode 
 \centerline{\epsfig{file=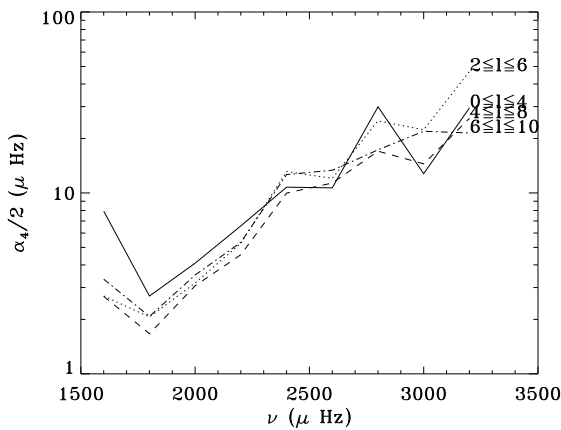,width=8.0cm}}
  \end{center} 
  \caption{\em Asymmetry parameter (separation between troughs and nulls) for the same
  modes as in Figure 3. } 
%  \label{fig:residuals_S} 
\end{figure} 

\subsection{Comparison with Theory}

Numerical peak and trough frequencies are calculated as follows. Eigenmode
frequencies for a standard solar model are determined in the usual way by 
solving the oscillation equations with the appropriate causal upper boundary
condition. The model is then truncated at some interior mesh point, corresponding to
the source location, and the
oscillation equations are solved with the upper boundary condition $\delta p=0$ applied at that point.
The frequencies obtained correspond to the nulls for the case $a_1=0$, which
\cite*{Rast+Bogdan98} refer to as a ``$\delta p$ source'' and which
\cite*{Rosenthal98} calls a monopole source. However it should be noted that this
is not the same usage of the term ``monopole'' as that of \cite*{Goldreich+Kumar90}.

The frequency differences between modes and troughs for various source depths are
converted into phase-function differences by multiplying by an empirically-determined
Duvall function. The measured peak-trough differences $\alpha_4/2$ are also converted
to phase-function differences by multiplying by the same Duvall function. Since each 
peak fitted is an average over several peaks, error estimates can be made by bootstrap resampling
of the peaks over which the averaging is performed. The results shown in Figure 5 are
broadly consistent with a fairly shallow source, located not more than about 250km 
below the solar surface, but evidently the results at this stage are somewhat preliminary. 
 \begin{figure}[!h] 
  \begin{center} 
    \leavevmode 
 \centerline{\epsfig{file=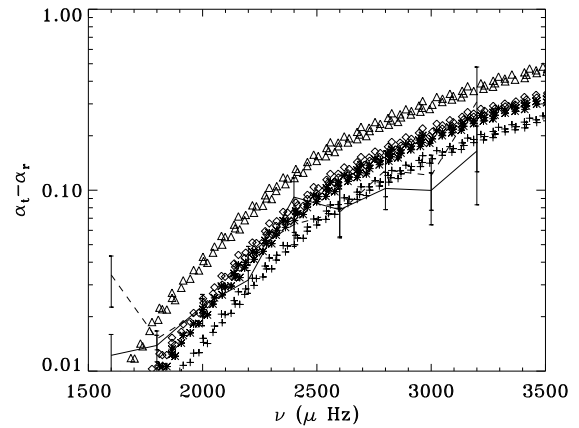,width=8.0cm}}
  \end{center} 
  \caption{\em Peak-trough separation calculated as a difference in phase function for
  modes in the range $0\!\le\!\ell\!\le\!10$. The solid curve is for prograde sectoral
  modes and the dashed curve is for retrograde sectoral modes. Error bars are 70\%
  confidence limits. Plotted symbols are numerical predictions for various source depths below
  the photosphere :- $+$-85km, $\ast$-200km, $\diamond$-250km and $\bigtriangleup$-500km. Note
  that 85km corresponds to the location of the sub-photospheric peak in the superadiabatic
  gradient in this model.} 
%  \label{fig:residuals_S} 
\end{figure} 

\subsection{Line Widths}

For completeness, Figure 6 shows the fitted line widths. The widths are
consistent with previously known results, particularly the dip in linewidth between 2.5mHz
and 2.8mHz, providing an extra validation
of the fitting procedure. The inclusion of asymmetry has no measurable effect
on the fitted widths. 
 \begin{figure}[!h] 
  \begin{center} 
    \leavevmode 
 \centerline{\epsfig{file=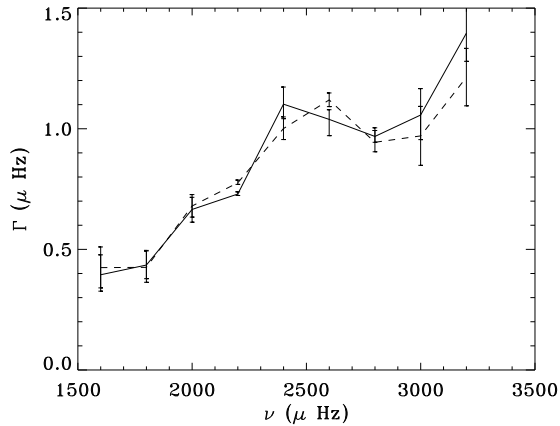,width=8.0cm}}
  \end{center} 
  \caption{\em Linewidths for prograde sectoral modes in the bin $0\!\le\!\ell\!\le\!10$. 
  The solid curve is for the asymmetric fit and the dashed curve is for the symmetric fit. 
  Error bars are 70\% confidence limits.} 
%  \label{fig:residuals_S} 
\end{figure} 

\section{Conclusions}

The general asymptotic theory of the excitation of solar oscillations 
demonstrates the relationship between the physics of the excitation and
the line-shape of the mode spectrum. Using this information, I have
shown how MDI data can be used to determine the form and location of
the excitation source. The re-analysis also provides corrections to the
mode frequencies. The consequences of these corrections for our understanding
of solar structure are discussed in more detail in \cite*{JCD+98}.

In the future, this analysis can be greatly refined
by the use of more data. In particular, by taking into account the m-leakage,
it will be possible to use information from whole multiplets, not just sectoral
modes as used here. This will allow much tighter limits to be placed on the
asymmetry parameter and its variation with frequency and degree. Crucially it
will also then be possible to determine any effect the asymmetry may have on the
determination of multiplet splittings.

We also need to consider
how to implement the fitting of these asymmetric profiles into the standard analysis
packages used to determine helioseismic frequencies from the data. 
An important issue is
likely to be the modelling of the background. A linearly sloping background could, 
in theory, 
mimic an asymmetric line profile, which is why I have here chosen a flat background. 
Standard fitting procedures assume backgrounds with various power-law forms and it
will be important to determine if the fitted parameters depend significantly on 
assumptions about the background (\cite{Kumar+Basu98}).

\section*{ACKNOWLEDGMENTS}

The National Center for Atmospheric Research is sponsored 
by the National Science Foundation. I acknowledge support from 
SOI/MDI NASA GRANT NAG5-3077. I am grateful to Mike Thompson
for first suggesting the use of the form (\ref{profile3}) and
to Jesper Schou for providing the data. 

\bibliographystyle{aabib}
\bibliography{aajour,convection,oscillations,Aake,books,jcd,Colin}

\end{document}